\documentclass[preprint,prl,10pt]{revtex4}%
\usepackage{amsfonts}
\usepackage{amsmath}
\usepackage{amssymb}
\usepackage{graphicx}%
\begin{document}
\preprint{cond-mat}
\title[ ]{Finite Size Effect on Correlation Functions of a Bose Gas in a Trap and \ \ \ \ \\Destruction of the Order Parameter by Phase Fluctuations. }
\author{V. S. Babichenko}
\affiliation{}
\author{}
\affiliation{}
\author{}
\affiliation{}
\pacs{PACS number}
\volumeyear{ }
\volumenumber{ }
\issuenumber{ }
\eid{ }
\date{15.05.2004}
\startpage{1}
\endpage{ }
\maketitle

An essential feature of \ ultra cold atomic gases, which are intensively
studied today, is there finite volume. From the point of view of the well
known Lee - Yang theorem the phase transitions in the finite size systems are
impossible \cite{YL}.\ An influence of finite sizes on the coherent properties
of a dilute Bose gas in a trap is analyzed in the present work. The trap is
described by the stationary external potential $U\left(  \overrightarrow
{r}\right)  $ which confines the system in a finite volume. The external
potential is supposed sufficiently\ smooth, so that the chemical potential
$\mu=ng$ is much larger than the distance between one-particle levels of the
external potential $U\left(  \overrightarrow{r}\right)  $, $n$\ is the average
density of the gas and $g$ is the coupling constant characterizing the
repulsive interaction between particles. Here and below we choose the system
of units in which the particle mass and Planck constant are equal to unit
$m=1$, $\hslash=1$.

The problem of the fluctuations of the condensate in the finite system with
the fixed total number of particles due to the long-wave phonon excitations
has been considered in the work \cite{GPS}. The anomalous dependence of these
fluctuations on the volume of the system was obtained there. The phase
fluctuations of the condensate in a highly elongated traps have been
considered in series of works, for example \cite{AKS}, \cite{L}. The mechanism
of fluctuations of the condensate in the finite Bose systems connected with
the fluctuations of the global phase has been considered in a series of works,
for example, \cite{LY}, \cite{CD}, \cite{G}. In these works the correlation
function $K_{\parallel}=<\widehat{\psi}\left(  \overrightarrow{r},t\right)
\widehat{\psi}^{+}\left(  \overrightarrow{r}^{\prime},t^{\prime}\right)  >$,
$\widehat{\psi}$ and $\widehat{\psi}^{+}$ being the annihilation and creation
operators of Bose particles, has been calculated and it has been shown that
this correlator decays exponentially for $\tau=\mid t-t^{\prime}%
\mid\longrightarrow\infty$ due to the phase diffusion. \ \ 

In the present work we consider the last mechanism of large fluctuations in
the finite Bose systems connected with the fluctuations of the global phase
and the existence of the zero mode, i.e., the zero energy excitation mode,
corresponding to the invariance of the action related to a change of the
global phase \cite{LY}. The method which we use differs from the methods of
the known works \cite{LY}, \cite{CD}, \cite{G} and gives the possibility to
consider quantum fluctuations and dissipation in the unified technique.

Due to existence of zero mode separated by the gap from the other excitation
spectrum, the phase fluctuations in finite systems have an infrared frequency
divergence independent of the space dimension. On the other hand, in the limit
of infinite sizes this infrared divergency disappears. As a result, the
properties of the correlation functions in the finite and infinite systems are
essentially different. Thus, the calculations of normal $K_{\parallel}$ and
anomalous $K_{\perp}=<\widehat{\psi}\left(  \overrightarrow{r},t\right)
\widehat{\psi}\left(  \overrightarrow{r}^{\prime},t^{\prime}\right)  >$
correlation functions at low temperatures show that correlator $K_{\parallel}$
decays exponentially in time for $\tau=\mid t-t^{\prime}\mid\longrightarrow
\infty$ and the anomalous correlator $K_{\perp}$ vanishes for all $\tau$.
\ The reason of such behavior of these correlators is the infrared frequency
divergency of phase fluctuations in the finite systems. At that, due to the
disappearance of this divergency in the case of the 3D system of infinite
sizes $L\longrightarrow\infty$,
$<$%
N%
$>$%
$\rightarrow\infty$, where L is the characteristic size of the system and
$<$%
N%
$>$
is the total number of particles, the normal and anomalous correlators restore
their usual properties, i.e., these correlators are equal to $n_{0}$ where
$n_{0}$ is the condensate density. Note that the anomalous averages are
considered in the sense of quasi-averaging, i.e., the infinitesimal term
$\widehat{j}^{+}\widehat{\psi}+\widehat{\psi}^{+}\widehat{j}$ breaking the
global gauge invariance is added to the Hamiltonian and the source
$\widehat{j}^{+}$ tends to zero at the end of the calculations. This
infinitesimal term results in the existence of the infinitesimal gap in the
energy of zero mode.\ The infinitesimal term will not be written evidently but
the existence of it will be implied always. The average $<\widehat{\psi
}\left(  \overrightarrow{r},t\right)  >$ , which usually considered as the
order parameter, is calculated too and being considered in the sense of
quasi-averaging gives the zero value. Note that all these averages are
calculated in the supposition of the thermodynamical equilibrium and are
accomplished over the equilibrium density matrix.

The action of a nonideal Bose gas in a stationary trap is%

\begin{equation}
S=\oint dt\int d^{d}r\left\{  \overline{\psi}\left[  i\partial_{t}+\mu
+\frac{1}{2}\overrightarrow{\nabla}^{2}-U\left(  \overrightarrow{r}\right)
-\frac{g}{2}\overline{\psi}\psi\right]  \psi\right\}  \tag{1}%
\end{equation}
The contour of the integration over the time is the two-time contour which was
introduced in the work \cite{Sch}. We use the Keldysh - Schwinger technique
\cite{Sch}, \cite{LVK} for the consideration of the system at low but finite
temperature and in the real time.

The anomalous and normal correlation functions can be written in the form of
the functional integral as%

\[
K_{\perp,\parallel}\left(  \overrightarrow{r}_{1},t_{1};\overrightarrow{r}%
_{2},t_{2}\right)  =\int\rho D\rho D\varphi\exp\left\{  iS\right\}
\rho\left(  x_{1}\right)  \rho\left(  x_{2}\right)  \exp\left\{  i\left(
\varphi\left(  x_{1}\right)  \pm\varphi\left(  x_{2}\right)  \right)
\right\}
\]

where we denote $x=\left(  \overrightarrow{r},t\right)  $, the complex field
$\psi$ is written in the modulus-phase representation $\psi=\rho e^{i\varphi}%
$, here and further the sings (+) and (-) relates to the anomalous and normal
correlators respectively. The under integration expression for the anomalous
and normal correlators can be represented as an exponent with the actions
$S_{J^{\left(  \perp\right)  }}$, $S_{J^{\left(  \parallel\right)  }}$ respectively%

\begin{equation}
K_{\perp,\parallel}\left(  \overrightarrow{r}_{1},t_{1};\overrightarrow{r}%
_{2},t_{2}\right)  =\int D\varphi Dn\rho\left(  x_{1}\right)  \rho\left(
x_{2}\right)  \exp\left\{  iS_{J^{\left(  \perp,\parallel\right)  }}\right\}
\tag{2}%
\end{equation}

where $S_{J^{\left(  \perp,\parallel\right)  }}$ are%

\begin{equation}
S_{J^{\left(  \perp,\parallel\right)  }}=\oint dt\int d^{d}r\left\{
J^{\left(  \perp,\parallel\right)  }\varphi+\overset{\cdot}{n}\varphi-\frac
{1}{2}n\left(  \overrightarrow{\nabla}\varphi\right)  ^{2}-\frac{1}{2}\left(
\overrightarrow{\nabla}\rho\right)  ^{2}-\frac{g}{2}n^{2}+n\left(
\mu-U\left(  \overrightarrow{r}\right)  \right)  \right\}  \tag{3}%
\end{equation}

Here we denote $n=\rho^{2}$, $\overset{\cdot}{\varphi}=\partial_{t}\varphi
$\ and
\[
J^{\left(  \perp,\parallel\right)  }\left(  \overrightarrow{r},t\right)
=\delta\left(  \overrightarrow{r}_{1}-\overrightarrow{r}\right)  \delta\left(
t-t_{1}\right)  \pm\delta\left(  \overrightarrow{r}_{2}-\overrightarrow
{r}\right)  \delta\left(  t-t_{2}\right)
\]

It is convenient to introduce new variables $\rho\rightarrow\rho$ and
$\varphi\rightarrow\xi=\rho\varphi$ that gives%

\begin{equation}
S=\oint dt\int d^{D}r\left\{  2\overset{\cdot}{\rho}\xi-\xi\widehat{H}_{\xi
}\xi-\frac{1}{2}\left(  \overrightarrow{\nabla}\rho\right)  ^{2}-\frac{g}%
{2}n^{2}+n\left(  \mu-U\left(  \overrightarrow{r}\right)  \right)  \right\}
\tag{4}%
\end{equation}
where%

\begin{equation}
\widehat{H}_{\xi}=-\frac{1}{2}\left(  \overrightarrow{\nabla}+\overrightarrow
{A}\right)  \left(  \overrightarrow{\nabla}-\overrightarrow{A}\right)
=\frac{1}{2}\left(  -\overrightarrow{\nabla}^{2}+\frac{\overrightarrow{\nabla
}^{2}\rho}{\rho}\right)  \tag{5}%
\end{equation}

and $\overrightarrow{A}=\frac{\overrightarrow{\nabla}\rho}{\rho}$.

Denote the orthonormal system of eigenfunctions and eigenvalues of the
Hamiltonian $\widehat{H}_{\xi}$ as $\eta_{\lambda}$ and $\varepsilon_{\xi
}\left(  \lambda\right)  $, correspondingly, $\lambda$ is a quantum number of
the eigenfunction\ $\eta_{\lambda}$,\ and%

\[
\widehat{H}_{\xi}\eta_{\lambda}=\varepsilon_{\xi}\left(  \lambda\right)
\eta_{\lambda}%
\]

The spectrum of the operator $\widehat{H}_{\xi}$ has the following properties:
firstly, it is a discontinuous one and $\lambda$ can be considered as an
integer, secondly, the eigenvalues of this operator have the zero or positive
value $\varepsilon_{\xi}\left(  \lambda\right)  \geqslant0$, thirdly, the zero
value $\varepsilon_{\xi}\left(  0\right)  =0$ exists for any configuration of
the field $\rho\left(  \overrightarrow{r},t\right)  $ and it is separated from
the other spectrum $\varepsilon_{\xi}\left(  \lambda\right)  $ $(\lambda
\neq0)$ by the gap which goes to zero when $L\rightarrow\infty$.

The system of eigenfunctions $\eta_{\lambda}$ can be chosen as an orthonormal system%

\[
<\eta_{\lambda^{\prime}}\mid\eta_{\lambda}>=\delta_{\lambda,\lambda^{\prime}}
\]

A wave function of the zero mode $\varepsilon_{\xi}\left(  0\right)  =0$ has
the form%

\begin{equation}
\eta_{0}\left(  \overrightarrow{r},t\right)  =\frac{1}{\sqrt{N}}\rho\left(
\overrightarrow{r},t\right)  \tag{6}%
\end{equation}

The existence of the zero mode can be seen directly from the action of the
operator (5) on the wave function (6). The eigenfunction expansion of $\xi$ is%

\[
\xi\left(  \overrightarrow{r},t\right)  =\sum\limits_{\lambda}\eta_{\lambda
}\left(  \overrightarrow{r},t\right)  a_{\lambda}\left(  t\right)
\]

and the zero mode in this expansion corresponds to the value $\lambda=0$.
Below we separate the zero mode in the eigenfunction expansion%

\begin{equation}
\xi\left(  \overrightarrow{r},t\right)  =\sum\limits_{\lambda}\eta_{\lambda
}\left(  \overrightarrow{r},t\right)  a_{\lambda}\left(  t\right)  =\eta
_{0}\left(  \overrightarrow{r},t\right)  a_{0}\left(  t\right)  +\xi^{\prime
}\left(  \overrightarrow{r},t\right)  \tag{7}%
\end{equation}

where%

\begin{equation}
\xi^{\prime}\left(  \overrightarrow{r},t\right)  =\sum\limits_{\lambda\neq
0}\eta_{\lambda}\left(  \overrightarrow{r},t\right)  a_{\lambda}\left(
t\right)  \tag{8}%
\end{equation}

As a result we obtain%

\begin{equation}
S=\oint dt\left\{  \frac{\overset{\cdot}{N}}{\sqrt{N}}a_{0}\left(  t\right)
\right\}  +\oint dt\int d^{d}r\left\{  2\overset{\cdot}{\rho}\xi^{\prime}%
-\xi^{\prime}\widehat{H}_{\xi}\xi^{\prime}-\frac{1}{2}\left(  \overrightarrow
{\nabla}\rho\right)  ^{2}-\frac{g}{2}n^{2}+n\left(  \mu-U\left(
\overrightarrow{r}\right)  \right)  \right\}  \tag{9}%
\end{equation}

Note, that the first term in the action S (9), as a result of the integration
over the zero mode $a_{0}\left(  t\right)  $ in the functional integral for
correlators (2), gives the conservation law of total number of particles ,
namely, the functional $\delta$-function $\delta\left[  \frac{\overset{\cdot
}{N}}{\sqrt{N}}\right]  $. This $\delta$-function gives the constraint for the
density fluctuations, and this constraint has to be taken into account when
the subsequent integration over the field $\rho$ is accomplished after the
integration over $a_{0}\left(  t\right)  $. Thereupon, the simplest way to
take into account the constraint connected with the conservation law of total
number of particles is the integration over $a_{0}\left(  t\right)  $ after
the integration over the density fluctuations, i.e., over the field $\rho$.

For the integration over the field $\rho$\ the saddle-point approximation can
be used. In this approximation the value of the field $\rho$ should have the
most symmetrical form and, thus, does not depend on the time t. The field
$\rho$ is represented as $\rho\left(  \overrightarrow{r},t\right)  =\rho
_{0}\left(  \overrightarrow{r}\right)  +\delta\rho\left(  \overrightarrow
{r},t\right)  $, where $\rho_{0}\left(  \overrightarrow{r}\right)  $ is the
saddle-point configuration and $\delta\rho\left(  \overrightarrow{r},t\right)
$ is the fluctuation. The saddle-point configuration $\rho_{0}\left(
\overrightarrow{r}\right)  $ is defined by the Eq.%

\[
\left[  -\frac{1}{2}\overrightarrow{\nabla}^{2}-\mu+U\left(  \overrightarrow
{r}\right)  +g\rho_{0}^{2}\right]  \rho_{0}=0
\]

The variation of $S$ can be considered up to the second order of fluctuations
$\delta\rho$ and can be represented in the form%

\begin{equation}
\delta S=\oint dt\left\{  a_{0}\left(  t\right)  \frac{\overset{\cdot}{\delta
N}}{\sqrt{N_{0}}}\right\}  +\oint dt\int d^{d}r\left\{  2\overset{\cdot
}{\delta\rho}\xi^{\prime}-\xi^{\prime}\widehat{H}_{\xi}^{\left(  0\right)
}\xi^{\prime}-\delta\rho\widehat{H}_{\rho}^{\left(  0\right)  }\delta
\rho\right\}  \tag{10}%
\end{equation}

where%

\[
\widehat{H}_{\xi}^{\left(  0\right)  }=\frac{1}{2}\left(  -\overrightarrow
{\nabla}^{2}+\frac{\overrightarrow{\nabla}^{2}\rho_{0}}{\rho_{0}}\right)
\]

and%

\[
\widehat{H}_{\rho}^{\left(  0\right)  }=-\frac{1}{2}\overrightarrow{\nabla
}^{2}-\mu+3g\rho_{0}^{2}+U\left(  \overrightarrow{r}\right)
\]

The value $\delta N$ is $\delta N\left(  t\right)  =\int d^{d}r2\rho
_{0}\left(  \overrightarrow{r}\right)  \delta\rho\left(  \overrightarrow
{r},t\right)  $ and $N_{0}=\int d^{d}r\rho_{0}^{2}\left(  \overrightarrow
{r}\right)  $.

The integration over $\delta\rho$ in the Eqs. (2) for the correlators
$K_{\perp}$ and $K_{\parallel}$ gives%

\[
K_{\perp,\parallel}\left(  \overrightarrow{r}_{1},\overrightarrow{r}_{2}%
;t_{1},t_{2}\right)  =\rho_{0}\left(  \overrightarrow{r}_{1}\right)  \rho
_{0}\left(  \overrightarrow{r}_{2}\right)  \int D\xi\exp\left\{  \delta
S_{J}\left[  \xi\right]  \right\}
\]

where%

\begin{equation}
\delta S_{J}\left[  \xi\right]  =\oint dt\int d^{d}r\left\{
\begin{array}
[c]{c}%
\left(  \frac{1}{\sqrt{N_{0}}}\rho_{0}\dot{a}_{0}+\dot{\xi}^{\prime}\right)
\frac{1}{\widehat{H}_{\rho}^{\left(  0\right)  }}\left(  \frac{1}{\sqrt{N_{0}%
}}\rho_{0}\dot{a}_{0}+\dot{\xi}^{\prime}\right)  -\xi^{\prime}\widehat{H}%
_{\xi}^{\left(  0\right)  }\xi^{\prime}+\\
+J\left(  \frac{1}{2\rho_{0}}\xi^{\prime}+\frac{1}{\sqrt{N_{0}}}a_{0}\right)
\end{array}
\right\}  \tag{11}%
\end{equation}

As it will be seen later the main contribution to the functional integral over
the field $a_{0}\left(  t\right)  $ gives the low frequency configurations,
such that their frequencies are much smaller than the frequencies of the field
$\xi^{\prime}$. Due to this fact the terms proportional to $\dot{a}_{0}%
\dot{\xi}^{\prime}$\ can be omitted in the action (11). Moreover, taking into
account only the slow configurations of the field $\xi^{\prime}$ which have
the wave vectors smaller than $\sqrt{\mu}$\ we can replace the term $\left(
\widehat{H}_{\rho}^{\left(  0\right)  }\right)  ^{-1}$\ by $\frac{1}{2\mu
}\delta\left(  \overrightarrow{r}-\overrightarrow{r}^{\prime}\right)  $ and,
thereby, rewrite the action (11) as%

\begin{equation}
\delta S_{J}\left[  \xi\right]  =\oint dt\left\{  \frac{1}{2\mu}\dot{a}%
_{0}^{2}\right\}  +\oint dt\int d^{d}r\left\{  \frac{1}{2\mu}\left(  \dot{\xi
}^{\prime}\right)  ^{2}-\xi^{\prime}\widehat{H}_{\xi}^{\left(  0\right)  }%
\xi^{\prime}+J\left(  \frac{1}{2\rho_{0}}\xi^{\prime}+\frac{1}{\sqrt{N_{0}}%
}a_{0}\right)  \right\}  \tag{12}%
\end{equation}

The integration over $a_{0}\left(  t\right)  $ and $\xi^{\prime}\left(
\overrightarrow{r},t\right)  $ gives the following result for the correlators%

\begin{equation}
K_{\perp,\parallel}\left(  \overrightarrow{r}_{1},\overrightarrow{r}_{2}%
;t_{1},t_{2}\right)  =\rho_{0}\left(  \overrightarrow{r}_{1}\right)  \rho
_{0}\left(  \overrightarrow{r}_{2}\right)  \exp\left\{  iS_{\perp,\parallel
}\right\}  \tag{13}%
\end{equation}

\begin{align*}
S_{\perp,\parallel}  &  =\frac{-1}{N_{0}}%
{\displaystyle\int}
\frac{d\omega}{2\pi}\left[  D_{11}^{\left(  0\right)  }\left(  \omega\right)
+D_{22}^{\left(  0\right)  }\left(  \omega\right)  \pm\left(  D_{12}^{\left(
0\right)  }\left(  \omega\right)  +D_{21}^{\left(  0\right)  }\left(
\omega\right)  \right)  \cos\left(  \omega\tau\right)  \right]  -\\
&  -%
{\displaystyle\int}
\frac{d\omega}{2\pi}\left\{  D_{11}^{\prime}\left(  \overrightarrow{r}%
_{1},\overrightarrow{r}_{2};\omega\right)  +D_{22}^{\prime}\left(
\overrightarrow{r}_{1},\overrightarrow{r}_{2};\omega\right)  \pm\left[
D_{12}^{\prime}\left(  \overrightarrow{r}_{1},\overrightarrow{r}_{2}%
;\omega\right)  +D_{21}^{\prime}\left(  \overrightarrow{r}_{1},\overrightarrow
{r}_{2};\omega\right)  \right]  \cos\left(  \omega\tau\right)  \right\}
\end{align*}
where $D_{\alpha,\beta}^{\left(  0\right)  }\left(  \omega\right)  $ is the
Keldysh-Schwinger Green function in "four-angle" representation \cite{LVK} for
the field $a_{0}\left(  t\right)  $ and $D_{\alpha,\beta}^{\prime}\left(
\overrightarrow{r}_{1},\overrightarrow{r}_{2};\omega\right)  $ is the
Keldysh-Schwinger Green function for the field $\xi^{\prime}\left(
\overrightarrow{r},t\right)  $, indices $\alpha,\beta=1,2$ means the upper and
the lower branches of the Keldysh-Schwinger time contour. As before, the upper
and the lower signs relate to $K_{\perp}$ and $K_{\parallel}$ respectively.
Eq. (14) can be rewritten in the form%

\begin{equation}
S_{\perp,\parallel}==\frac{-2}{N_{0}}%
{\displaystyle\int}
\frac{d\omega}{2\pi}D_{K}^{\left(  0\right)  }\left(  \omega\right)  \left(
1\pm\cos\left(  \omega\tau\right)  \right)  -2%
{\displaystyle\int}
\frac{d\omega}{2\pi}%
{\displaystyle\sum\limits_{\lambda\neq0}}
D_{K}^{\prime}\left(  \lambda;\omega\right)  \eta_{\lambda}\left(
\overrightarrow{r}_{1}\right)  \overline{\eta}_{\lambda}\left(
\overrightarrow{r}_{2}\right)  \left(  1\pm\cos\left(  \omega\tau\right)
\right)  \tag{14}%
\end{equation}

The Eq. (14) is obtained from the Eq. for $S_{\perp,\parallel}$ (13) using the
following property of Keldysh-Schwinger Green functions in "four-angle" representation%

\[
D_{11}\left(  \omega\right)  +D_{22}\left(  \omega\right)  -D_{12}\left(
\omega\right)  -D_{21}\left(  \omega\right)  =0
\]
and the following connections between non-diagonal components of these Green
functions in "four-angle" representation, "kinetic" Green function $D_{K}$, as
well as the retarded $D_{R}$ and advanced $D_{A}$ Green functions%

\[
D_{12}\left(  \omega\right)  +D_{21}\left(  \omega\right)  =2D_{K}\left(
\omega\right)  =2\coth\left(  \frac{\omega}{2T}\right)  \left[  D_{R}\left(
\omega\right)  -D_{A}\left(  \omega\right)  \right]
\]
The last Eqs. take place in equilibrium states. If the relaxation processes
are not taken into account the Green functions $D_{R,A}\left(  \omega\right)
$\ and $D_{K}^{\left(  0\right)  }\left(  \omega\right)  $\ have the form%

\begin{equation}
D_{R,A}^{\left(  0\right)  }\left(  \omega\right)  =\frac{2\mu}{\omega^{2}\pm
i\delta};\text{ }D_{R,A}^{\prime}\left(  \lambda;\omega\right)  =\frac{2\mu
}{\omega^{2}-2\mu\varepsilon_{\xi}\left(  \lambda\right)  \pm i\delta};\text{
}D_{K}^{\left(  0\right)  }\left(  \omega\right)  =2\pi i\coth\left(
\frac{\omega}{2T}\right)  sign\left(  \omega\right)  \mu\delta\left[
\omega^{2}\right]  \tag{15}%
\end{equation}

The regularization of the first integral in the Eq. (14) by means of the
change $\delta\left[  \omega^{2}\right]  \rightarrow$ $\delta\left[
\omega^{2}-\varepsilon_{0}^{2}\right]  $ and then the transition to the limit
$\varepsilon_{0}\rightarrow0$ results in \ \ \ \ \ \ \ \
\[
S_{\perp}=i\lim_{\varepsilon_{0}\rightarrow0}\left(  \frac{2}{N_{0}}\frac
{T\mu}{\varepsilon_{0}^{2}}\right)  ;\text{ \ \ \ \ \ \ \ \ \ \ \ }%
S_{\parallel}=i\frac{2}{N_{0}}T\mu\tau^{2}%
\]
, at that, the first integral in the Eq. (14) gives the main contribution to
$S_{\perp,\parallel}$, thus, it gives the infinite contribution to $S_{\perp}%
$, and the second integral gives the finite contribution to $S_{\perp}$ due to
the existence of the gap in the spectrum of the $\xi^{\prime}$ excitations.
Obviously, that these Eqs. result in the zero value for the anomalous
correlator $K_{\perp}\left(  \tau\right)  =0$ and the exponential decrease of
the normal correlator $K_{\parallel}\left(  \tau\right)  =\rho_{0}^{2}%
\exp\left(  -\frac{4}{N_{0}}T\mu\tau^{2}\right)  $ for the sufficiently large
$\tau$.

Note, that in the limit of the infinite size of the system L$\rightarrow
\infty$ the values $S_{\perp}$, $S_{\parallel}$ transfer to%

\begin{equation}
S_{\perp,\parallel}=i\frac{2}{n_{0}}%
{\displaystyle\int}
\frac{d\omega d^{d}k}{\left(  2\pi\right)  ^{d+1}}\left[  D_{K}\left(
\omega,\overrightarrow{k}\right)  \left(  1\pm\cos\left(  \omega\tau\right)
\right)  \right]  \tag{16}%
\end{equation}

where%

\[
D_{K}\left(  \omega,\overrightarrow{k}\right)  =2\pi i\coth\left(
\frac{\omega}{2T}\right)  sign\left(  \omega\right)  \mu\delta\left[
\omega^{2}-c^{2}k^{2}\right]
\]
In this case due to the integration over the momentum $\overrightarrow{k}$
there is no divergency of the integral over the small frequencies and with an
accuracy to the small gas parameter in the dilute gas the anomalous and normal
correlators takes the usual form $K_{\parallel}=K_{\perp}=\rho_{0}^{2}$ in 3D
case. This fact can be obtained easily by the calculation of the integral
(16), more over, this calculation gives the well known vanish of $K_{\perp}$
in 2D case for a nonzero temperature and in 1D case for any temperature.

The singularity of the expression for $D_{K}^{\left(  0\right)  }\left(
\omega\right)  $\ strongly changes by the relaxation of the phase
fluctuations. To obtain the action taking into account the relaxation, the
terms proportional to the fourth power of $\delta\rho$, i.e., $g\left(
\delta\rho\right)  ^{4}$ should be considered. Integrating over the rapid
fields $\delta\rho$ and $\xi^{\prime}$, i.e., the fields having the momentums
$k\gtrsim$ $T/c$, where c is the sound velocity and the temperature T is
supposed being much smaller than $\mu$, we obtain the additional term
$\delta\rho\Sigma\delta\rho$ in the effective action for the slow fields. In
the case of the dilute Bose gas the self-energy part $\Sigma$ is determined by
the diagram of the second order in the coupling constant $g$ and has three
internal particle lines corresponding to Green functions of the rapid
$\delta\rho$\ field $G_{\delta\rho,\delta\rho}=-i<T_{t}\widehat{\delta\rho
}\left(  x\right)  \widehat{\delta\rho}\left(  x^{\prime}\right)  >$, where
$T_{t}$ means t-ordering of operators. The additional self-energy part for the
slow $\delta\rho$\ field can be written as $\Sigma\left(  x_{1},x_{2}\right)
=g^{2}\left(  G_{\delta\rho,\delta\rho}\left(  x_{1},x_{2}\right)  \right)
^{3}$ and, namely, this term describes the relaxation processes. In the case
of the dilute Bose gas only the imaginary part of this self-energy part can be
taken into account, the real part can be omitted due to its smallness compared
with the chemical potential. The self-energy part $\Sigma$ for small
frequencies $\omega\rightarrow0$\ is calculated as $\Sigma_{R,A}\sim
\frac{g^{2}}{\mu^{6}}\omega T^{7}$ and the kinetic part $\Sigma_{K}$
\cite{LVK}, \cite{Sch}\ takes the form $\Sigma_{K}\sim\frac{g^{2}}{\mu^{6}%
}T^{8}$. Note that the internal momentums which give the main contribution to
the diagram for $\Sigma$ are $p_{1}\sim p_{2}\sim T/c$. After the integration
over $\delta\rho$\ and taking into account the additional term $\delta
\rho\Sigma\delta\rho$ the function $\Gamma\left(  \omega\right)  $ is
calculated as $\Gamma\left(  \omega\right)  \sim\frac{g^{2}}{\mu^{8}}%
\omega^{3}T^{7}$.

After the transformation to the action being written in terms of \ slow fields
$\psi$ and containing the relaxation term the transition to the modulus-phase
variables for the slow fields can be made. After that the subsequent scheme of
the transformations is analogous to those which has been made above, i.e.,
without consideration of the relaxation term. This gives the expressions for
$S_{\perp,\parallel}$ in the form (14) where the phase correlators
$D_{R,A}^{\left(  0\right)  }\left(  \omega\right)  $ and $D_{R,A}^{^{\prime}%
}\left(  \omega,\lambda\right)  $ have to be changed by%

\begin{equation}
D_{R,A}^{\left(  0\right)  }\left(  \omega\right)  =\frac{1}{\omega^{2}\pm
i\Gamma\left(  \omega\right)  };\text{ \ \ \ \ \ \ }D_{R,A}^{^{\prime}}\left(
\omega,\lambda\right)  =\frac{1}{\omega^{2}-2\mu\varepsilon_{\xi}\left(
\lambda\right)  \pm i\Gamma\left(  \omega\right)  } \tag{17}%
\end{equation}

In the case of the low temperatures $T<<\mu$ and for small frequencies
$\omega\rightarrow0$ the relaxation $\Gamma\left(  \omega\right)  $ can be
calculated as $\Gamma\left(  \omega\right)  \sim\omega\frac{\left(
T/c\right)  ^{6}}{T^{2}}g^{2}$, where c is the sound velocity. Note, that the
internal momentums which give the main contribution to the diagram for the
relaxation $\Gamma\left(  \omega\right)  $ are $p_{1}\sim p_{2}\sim T/c$. At
that, the kinetic Green function has the form
\begin{equation}
D_{K}^{\left(  0\right)  }\left(  \omega\right)  =-i\coth\left(  \frac{\omega
}{2T}\right)  \frac{\Gamma\left(  \omega\right)  }{\omega^{4}+\Gamma
^{2}\left(  \omega\right)  };\text{ \ }D_{K}^{^{\prime}}\left(  \omega
,\lambda\right)  =-i\coth\left(  \frac{\omega}{2T}\right)  \frac{\Gamma\left(
\omega\right)  }{\left(  \omega^{2}-2\mu\varepsilon_{\xi}\left(
\lambda\right)  \right)  ^{2}+\Gamma^{2}\left(  \omega\right)  }\text{\ \ \ }
\tag{18}%
\end{equation}

The kinetic Green function $D_{K}^{\left(  0\right)  }\left(  \omega\right)  $
for small frequencies $\omega\rightarrow0$ can be written as $D_{K}^{\left(
0\right)  }\left(  \omega\right)  =\frac{-i\mu^{2}\Gamma\left(  \omega\right)
}{\omega^{4}}\coth\left(  \frac{\omega}{2T}\right)  \sim-i\frac{1}{\mu
^{6}\omega^{2}}g^{2}T^{8}$, i.e., it has the nonintegrable singularity for
$\omega\rightarrow0$, in contrast to the correlator $D_{K}^{^{\prime}}\left(
\omega,\lambda\right)  $. Thus, the integral for $S_{\perp}$ (14) diverges at
small frequencies as $1/\omega$. Due to the divergency of $S_{\perp}$\ the
anomalous correlation function $K_{\perp}$ vanishes $K_{\perp}\left(
\tau\right)  =0$. The normal correlator $K_{\parallel}\left(  \tau\right)  $
for sufficiently large $\tau$ decays exponentially and has the form
$K_{\parallel}\left(  \tau\right)  =\rho_{0}^{2}\exp\left(  -\frac{1}{N_{0}%
}2\mu^{2}T\left[  \frac{\Gamma\left(  \omega\right)  }{\omega^{3}}\right]
_{\omega\rightarrow0}\mid\tau\mid\right)  $. Note that nonzero value of the
relaxation $\Gamma\left(  \omega\right)  $ for small frequencies is obtained
only in the case when the deviation of the excitation spectrum from the linear
one for small momentums has the positive sign, and when this sign is negative
the relaxation $\Gamma\left(  \omega\right)  $ has zero value.

In conclusion, we note that the calculation of the order parameter
$<\widehat{\psi}>$ is analogous to the calculation of $K_{\perp}$\ and gives
the lack of the order parameter $<\widehat{\psi}>=0$ in the finite Bose
systems as well as vanishing of $K_{\perp}$. On the other hand, in the 3D case
in the limit L$\rightarrow\infty$\ the anomalous correlator $K_{\perp}$ as
well as the normal correlator $K_{\parallel}$ and the order parameter
$<\widehat{\psi}>$\ recover their usual values $K_{\perp}=K_{\parallel
}=\left(  <\widehat{\psi}>\right)  ^{2}=\rho_{0}^{2}$.

This work was supported by the Russian Foundation for Basic Research, by the
Netherlands Organization for Scientific Research (NWO) and by INTAS -2001-2344.

\bigskip

\section{}
\end{document}